\documentstyle[pra,aps,multicol]{revtex}
\renewcommand{\narrowtext}{\begin{multicols}{2} \global\columnwidth18pc}
\renewcommand{\widetext}{\end{multicols} \global\columnwidth38pc}
\begin{document}
\draft
\title{{\bf A necessary and sufficient criterion for multipartite
separable states}
}
\author{Shengjun Wu}
\address{{\it
Department of Modern Physics,
University of Science and Technology of China, Hefei 230027, P.R.China  \\
}}
\author{Xuemei Chen \footnote{Email: xmchen@ustc.edu.cn} }
\address{{\it
Department of Physics,
University of Science and Technology of China, Hefei 230026, P.R.China  \\
}}
\author{Yongde Zhang \footnote{Email: ydzhang@ustc.edu.cn} }
\address{{\it
CCAST(WORLD LABORATORY) P.O.BOX 8730, Beijing 100080 and \\        
Department of Modern Physics,
University of Science and Technology of China, Hefei 230027, P.R.China  
}}

\date{\today}
\maketitle

\noindent \hrulefill
%\begin{abstract}

\noindent {\bf Abstract}

\vskip 0.3cm

We present a necessary and sufficient condition for the separability of
multipartite quantum states, this criterion also tells us how to write a
multipartite separable state as a convex sum of separable pure states. To
work out this criterion, we need to solve a set of equations, actually it
is easy to solve these equations analytically if the density matrix of the
given quantum state has few nonzero eigenvalues.

\vskip 0.3cm



%\end{abstract}
\noindent \hrulefill

\vskip 1.0cm

%\narrowtext

Ever since it was first noted by Einstein-Podolsky-Rosen (EPR) \cite{epr}
and Schr\"odinger \cite{Sc}, entanglement has played an important role in
quantum information theory. Quantum entanglement provides strong tests of
quantum nonlocality \cite{Be,bc}, and it is also a useful resource for
various kinds of quantum information processing, including teleportation 
\cite{bbcjpw,bpz}, entanglement swapping \cite{zzhe,pbwz}, cryptographic key distribution \cite{des}, quantum error
correction \cite{Sh} and quantum computation \cite{De}.

A multipartite quantum state is called separable if it can be written as
a convex sum of product states belonging to different parties,
otherwise it is called entangled.
It is important to know whether a given multipartite quantum state is separable or entangled.

So far, there have been many ingenious separability criteria. Since a
separable state always satisfies Bell's inequalities, the latter
represent a necessary condition for separability \cite{We}, but generally 
they are not sufficient. Peres \cite{Pe}
discovered another simple necessary condition for separability, a partial
transposition of a bipartite quantum state $\rho _{AB}$ with respect to a
subsystem $A$ (or $B$) must be positive if $\rho _{AB}$ is separable. Peres'
criterion has been shown by Horodecki et al. \cite{MPR} to be strong enough
to guarantee separability for bipartite systems of dimension $2\times 2$ or $%
2\times 3$, but, for other cases it is not a sufficient one.
It has been proved by Horodecki et al. \cite{MPR} that a necessary and
sufficient condition for separability of bipartite mixed state is its
positivity under all the maps of the form $I\otimes \Lambda$, where
$\Lambda$ is any positive map. This criterion is
more important in theory than in practice since it involves the
characterization of the set of all positive maps, which is not easy.
More recently, Horodecki-Horodecki\cite{MP} and Cerf-Adami-Gingrich \cite{cag} have
independently derived a reduction criterion of separability for bipartite
quantum states, this criterion is equivalent to Peres' for $2\times n$
composite systems, and it is not sufficient for separability in
general cases. Many interesting separability criteria have been presented
recently, such as the rank separability criterion derived by Horodecki et al.
\cite{hstt}, which shows that a separable state cannot have the rank of a
reduced density matrix greater than the rank of total density matrix, this
necessary condition is easy for operation. 

Here we introduce a necessary and sufficient condition for the separability of
multipartite quantum states, this criterion also gives the expression for
a separable state in the form of convex sum of product pure states.

Let there be $m$ subsystems A, B, $\cdots $, M belonging to $m$
different observers Alice, Bob, $\cdots $, Mary, respectively. A m-party quantum state $%
\rho _{AB\cdots M}$ is called separable iff it can be written as 
\begin{eqnarray}
\rho _{AB\cdots M} = \sum_{i=1}^rp_i\left| \psi _i^A\psi _i^B\cdots \psi
_i^M\right\rangle \left\langle \psi _i^A\psi _i^B\cdots \psi _i^M\right|
\label{wu1}
\end{eqnarray}
where $\left\{ \left| \psi _i^\alpha \right\rangle \left| i=1,2,\cdots
r\right. \right\} $ is a set of normalized (generally not orthogonal) states
of system $\alpha $ ($\alpha =$A,B,$\cdots $,M), and the probabilities $%
p_i>0 $, $\sum_i^rp_i=1$. On the other hand, any given quantum state (no
matter it is separable or entangled) $\rho _{AB\cdots M}$ can always be
written in the orthogonal representation as 
\begin{equation}
\rho _{AB\cdots M}=\sum_{i=1}^k\lambda _i\left| \phi _i^{AB\cdots
M}\right\rangle \left\langle \phi _i^{AB\cdots M}\right|  \label{wu2}
\end{equation}
where $\left| \phi _i^{AB\cdots M}\right\rangle $ is a set of normalized
orthogonal eigenstates corresponding to the nonzero eigenvalues $\lambda _i$(%
$\lambda _i>0$, $\sum_i^k\lambda _i=1$). The eigenstates and eigenvalues of $%
\rho _{AB\cdots M}$ can always be solved by a standard procedure.

\textbf{Theorem:} Let $\left| \phi _i^{AB\cdots M}\right\rangle $ be the eigenstates
corresponding to the nonzero eigenvalues $\lambda _i$ ($i=1,\cdots k$)
for a given m-party quantum state $\rho _{AB\cdots M}$,
$\rho _{AB\cdots M}$ is separable if and only if the equations
\begin{equation}
\left\{ 
\begin{array}{l}
\left| \Psi \right\rangle \equiv \sum_{i=1}^ky_i\left| \phi _i^{AB\cdots
M}\right\rangle \\
|y_1|^2+|y_2|^2+\cdots +|y_k|^2=1 \\ 
\sigma_\alpha \equiv tr_{\overline{\alpha}} \left(\left| \Psi \right\rangle
\left\langle \Psi \right| \right)    \\
\det \left( \sigma _\alpha -I\right) =0 \;\; \left( \alpha =A,B,\cdots M\right)
\end{array}
\right.  \label{wuth}
\end{equation}
(here $\alpha$ denotes one of the $m$ parties, and $\overline{\alpha}$ denotes
the remaining $m-1$ parties) have $r$ different vector solutions
$\overrightarrow{y}^{\left( l\right) }$ ($%
l=1,2,\cdots ,r;r\geq k$) satisfying the following condition: there exists a
set of positive numbers $p_i$ ($\sum_i^rp_i=1$), so that 
\begin{equation}
\left\{ 
\begin{array}{c}
M_{ij}\equiv \sqrt{\frac{p_i}{\lambda _j}}y_j^{\left( i\right) } \\ 
M^{\dagger }M=I_{k\times k}
\end{array}
\right.  \label{wu3}
\end{equation}
Moreover, if $\rho _{AB\cdots M}$ is separable, it can be written as the
following mixture of separable pure states 
\begin{equation}
\rho _{AB\cdots M}=\sum_{i=1}^rp_i\left| \psi _i^A\psi _i^B\cdots \psi
_i^M\right\rangle \left\langle \psi _i^A\psi _i^B\cdots \psi _i^M\right| 
\end{equation}
where $p_i$ is given by Eq. (\ref{wu3}), and $\left| \psi _i^A\psi
_i^B\cdots \psi _i^M\right\rangle $ is given by 
\begin{equation}
\left| \psi _i^A\psi _i^B\cdots \psi _i^M\right\rangle =\sum_jy_j^{\left(
i\right) }\left| \phi _j^{AB\cdots M}\right\rangle    \label{wuth1}
\end{equation}

Here, we say two vectors $\overrightarrow{y}^{\left( 1\right) }$, $%
\overrightarrow{y}^{\left( 2\right) }$ are different if there exists no
factor $K$ such that $\overrightarrow{y}^{\left( 2\right) }=K\cdot 
\overrightarrow{y}^{\left( 1\right) }$. Actually we can always choose the
first column of the matrix $M$ (i.e., $y_1^{\left( l\right) }$) to be
non-negative real numbers. And obviously, there are only $m-1$
independent equations among the $m$ equations $\det
\left( \sigma _\alpha -I\right) =0$ $\left( \alpha =A,B,\cdots M\right) $.

Proof. The theorem can be proved using Hughston-Jozsa-Wootters'
result \cite{hjw} and the properties of the separable pure states,
while in the following, we give a simple proof of the theorem directly.

Let us first prove the necessity. Suppose the state $\rho _{AB\cdots M}$
is separable, i.e.,
\begin{equation}
\rho _{AB\cdots M}=\sum_{i=1}^rp_i\left| \psi _i^A\psi _i^B\cdots \psi
_i^M\right\rangle \left\langle \psi _i^A\psi _i^B\cdots \psi _i^M\right| 
\end{equation}
set $y_j^{\left( i\right) }=\left\langle \phi _j^{AB\cdots M}\right. \left|
\psi _i^A\psi _i^B\cdots \psi _i^M\right\rangle $. It is obvious that $%
\overrightarrow{y}^{\left( i\right) }$ is the $i$-th ($i=1,\cdots ,r$)
solution of Eqs. (\ref{wuth}), since the state 
\begin{equation}
\left| \Psi ^{\left( i\right) }\right\rangle \equiv \sum_{j=1}^ky_j^{\left(
i\right) }\left| \phi _j^{AB\cdots M}\right\rangle =\left| \psi _i^A\psi
_i^B\cdots \psi _i^M\right\rangle 
\end{equation}
is a separable pure state. Set $M_{ij}\equiv \sqrt{\frac{p_i}{\lambda _j}}%
y_j^{\left( i\right) }$, we can easily to show that $M^{\dagger }M=I_{k\times
k} $ since  
\begin{eqnarray}
\rho _{AB\cdots M}
&=&\sum_{i=1}^rp_i\left| \psi _i^A\psi _i^B\cdots \psi _i^M\right\rangle
\left\langle \psi _i^A\psi _i^B\cdots \psi _i^M\right|  \nonumber \\
&=&\sum_{i=1}^r\sum_{j,j^{\prime }=1}^kp_iy_j^{\left( i\right) }y_{j^{\prime
}}^{\left( i\right) *}\left| \phi _j^{AB\cdots M}\right\rangle \left\langle
\phi _{j^{\prime }}^{AB\cdots M}\right|  \nonumber  \\
&=&\sum_{j,j^{\prime }=1}^k\sqrt{\lambda _j\lambda _{j^{\prime }}}\left|
\phi _j^{AB\cdots M}\right\rangle \left\langle \phi _{j^{\prime }}^{AB\cdots
M}\right| \cdot \sum_{i=1}^rM_{ij}M_{ij^{\prime }}^{*}
\end{eqnarray}
and
\begin{eqnarray}
\rho _{AB\cdots M} &=&\sum_{i=1}^k\lambda _i\left| \phi _i^{AB\cdots
M}\right\rangle \left\langle \phi _i^{AB\cdots M}\right|  \nonumber \\
&=&\sum_{j,j^{\prime }=1}^k\sqrt{\lambda _j\lambda _{j^{\prime }}}\left|
\phi _j^{AB\cdots M}\right\rangle \left\langle \phi _{j^{\prime }}^{AB\cdots
M}\right| \cdot \delta_{jj^{\prime}}
\end{eqnarray}
This completes the proof of necessity.

Next we come to prove the sufficiency. Suppose the Eqs. (\ref{wuth}) have
already had solutions $\overrightarrow{y}^{\left( l\right) }$ ($l=1,\cdots ,r;r\geq k$)
with a proper set of positive numbers $p_i$ satisfying Eqs. (\ref{wu3}), then the
state $\sum_{i=1}^ky_i^{\left( l\right) }\left| \phi _i^{AB\cdots
M}\right\rangle $ must be a separable pure state since $\det \left( \sigma
_\alpha -I\right) =0$ $\left( \alpha =A,B,\cdots M\right) $. Set $%
\sum_{i=1}^ky_i^{\left( l\right) }\left| \phi _i^{AB\cdots M}\right\rangle
=\left| \psi _l^A\psi _l^B\cdots \psi _l^M\right\rangle $. Now
we only need to show that
\begin{equation}
\rho _{AB\cdots M}=\sum_{l=1}^rp_l\left| \psi _l^A\psi _l^B\cdots \psi
_l^M\right\rangle \left\langle \psi _l^A\psi _l^B\cdots \psi _l^M\right| 
\end{equation}
This is obvious since we have 
\begin{eqnarray}
&&\sum_{l=1}^rp_l\left| \psi _l^A\psi _l^B\cdots \psi _l^M\right\rangle
\left\langle \psi _l^A\psi _l^B\cdots \psi _l^M\right|  \nonumber \\
&=&\sum_{l=1}^rp_l\cdot \sum_{i,j=1}^ky_i^{\left( l\right) }y_j^{\left(
l\right) *}\left| \phi _i^{AB\cdots M}\right\rangle \left\langle \phi
_j^{AB\cdots M}\right|  \nonumber \\
&=&\sum_{i,j=1}^k\sqrt{\lambda _i\lambda _j}\left| \phi _i^{AB\cdots
M}\right\rangle \left\langle \phi _j^{AB\cdots M}\right| \cdot
\sum_{l=1}^rM_{li}M_{lj}^{*} \nonumber  \\
&=&\sum_{i=1}^k\lambda _i\left| \phi _i^{AB\cdots M}\right\rangle
\left\langle \phi _i^{AB\cdots M}\right|  \nonumber  \\
&=&\rho _{AB\cdots M}
\end{eqnarray}
This completes the proof of sufficiency.

In the theorem, the separability of a given quantum state is determined
by solving a set of equations of the vector variable $\overrightarrow{y}=$($y_1$,
$y_2$, $\cdots 
$, $y_m$). If $\rho _{AB\cdots M}$ has few nonzero eigenvalues (i.e., $k$ is
small), generally we can get analytic solutions for Eqs. (\ref{wuth}). However, 
if $\rho _{AB\cdots M}$ has many nonzero eigenvalues (i.e., $k$ is great),
then it is difficult to work out analytic solutions for the equations in the
theorem, only numerical solutions are practical.

Here are some examples.

(1). Let 
\begin{equation}
\rho _{AB}=\lambda \left| \phi ^{+}\right\rangle \left\langle \phi
^{+}\right| +(1-\lambda )\left| \phi ^{-}\right\rangle \left\langle \phi
^{-}\right|  \label{wue1}
\end{equation}

As in the theorem, set 
\begin{eqnarray}
\left| \Psi \right\rangle &\equiv &y_1\left| \phi ^{+}\right\rangle
+y_2\left| \phi ^{-}\right\rangle  \nonumber  \\
&=&\frac{y_1+y_2}{\sqrt{2}}\left| 00\right\rangle +\frac{y_1-y_2}{\sqrt{2}}%
\left| 11\right\rangle
\end{eqnarray}
Direct calculation yields
\begin{equation}
\sigma _A=\frac{|y_1+y_2|^2}2\left| 0\right\rangle \left\langle 0\right| +%
\frac{|y_1-y_2|^2}2\left| 1\right\rangle \left\langle 1\right| 
\end{equation}
From $\det \left( \sigma _A-I\right) =0$, we get
\begin{equation}
y_1=\pm y_2 
\end{equation}
Considering the relation $|y_1|^2+|y_2|^2=1$, we have two (and only two)
different vector solutions:
\begin{eqnarray*}
\overrightarrow{y}^{\left( 1 \right)} &=&\left( \frac 1{\sqrt{2}},\frac 1{\sqrt{2}}\right) \\
\overrightarrow{y}^{\left( 2 \right)} &=&\left( \frac 1{\sqrt{2}},-\frac 1{\sqrt{2}}\right)
\end{eqnarray*}
So 
\begin{equation}
M=\left( 
\begin{array}{cc}
\frac 1{\sqrt{2}}\sqrt{\frac p\lambda } & \frac 1{\sqrt{2}}\sqrt{\frac
p{1-\lambda }} \\ 
\frac 1{\sqrt{2}}\sqrt{\frac{1-p}\lambda } & -\frac 1{\sqrt{2}}\sqrt{\frac{%
1-p}{1-\lambda }}
\end{array}
\right) 
\end{equation}
Let $M^{\dagger }M=I_{2\times 2}$, we have that 
\begin{eqnarray*}
\lambda &=&1-\lambda =\frac 12 \\
p &=&1-p=\frac 12
\end{eqnarray*}
And there is 
\begin{equation}
\left( 
\begin{array}{cc}
\frac 1{\sqrt{2}} & \frac 1{\sqrt{2}} \\ 
\frac 1{\sqrt{2}} & -\frac 1{\sqrt{2}}
\end{array}
\right) \left( 
\begin{array}{c}
\frac 1{\sqrt{2}}\left| \phi ^{+}\right\rangle \\ 
\frac 1{\sqrt{2}}\left| \phi ^{-}\right\rangle
\end{array}
\right) =\left( 
\begin{array}{c}
\frac 1{\sqrt{2}}\left| 00\right\rangle \\ 
\frac 1{\sqrt{2}}\left| 11\right\rangle
\end{array}
\right) 
\end{equation}
The conclusion is that for $\lambda =\frac 12$, $\rho _{AB}$ is
separable and $\rho _{AB}=\frac 12\left| 00\right\rangle \left\langle
00\right| +\frac 12\left| 11\right\rangle \left\langle 11\right| $, and for 
$\lambda \neq \frac 12$, $\rho _{AB}$ is entangled.

(2). Let
\begin{eqnarray}
\rho _{AB} &=&\frac 14\left[ \frac 1{\sqrt{2}}\left( \left| \phi
^{+}\right\rangle -i\left| \psi ^{+}\right\rangle \right) \right] \left[
\frac 1{\sqrt{2}}\left( \left\langle \phi ^{+}\right| +i\left\langle \psi
^{+}\right| \right) \right]  \nonumber \\
&&+\frac 34\left[ \frac 1{2\sqrt{3}}\left( -3i\left| 00\right\rangle
+i\left| 11\right\rangle +\left| 01\right\rangle +\left| 10\right\rangle
\right) \right] \nonumber \\
&&\cdot \left[ \frac 1{2\sqrt{3}}\left( 3i\left\langle 00\right|
-i\left\langle 11\right| +\left\langle 01\right| +\left\langle 10\right|
\right) \right]  \label{wue3}
\end{eqnarray}
Set 
\begin{eqnarray}
\left| \Psi \right\rangle &\equiv& y_1\cdot \frac 1{\sqrt{2}}\left( \left|
\phi ^{+}\right\rangle -i\left| \psi ^{+}\right\rangle \right)  \nonumber  \\
&&+y_2\cdot
\frac 1{2\sqrt{3}}\left( -3i\left| 00\right\rangle +i\left| 11\right\rangle
+\left| 01\right\rangle +\left| 10\right\rangle \right) 
\end{eqnarray}
For the convenience of calculation, denote $y_1=r_1$, $y_2=r_2\cdot
e^{i\varphi }$, here $r_1$, $r_2$ are positive numbers satisfying the
relation $r_1^2+r_2^2=1$, and $\varphi $ is a real number.

Direct calculation gives
\begin{eqnarray}
\sigma _A &=&\left( \frac 12+\frac 13r_2^2+\frac 1{\sqrt{3}}r_1r_2\sin \varphi
\right) \left| 0\right\rangle \left\langle 0\right|   \nonumber  \\
&&+\left( \frac 12-\frac
13r_2^2-\frac 1{\sqrt{3}}r_1r_2\sin \varphi \right) \left| 1\right\rangle
\left\langle 1\right|   \nonumber  \\
&&+\left( -\frac 13i\cdot r_2^2+\frac 1{\sqrt{3}}r_1r_2e^{i\varphi }\right)
\left| 0\right\rangle \left\langle 1\right|   \nonumber  \\
&&+\left( \frac 13i\cdot
r_2^2+\frac 1{\sqrt{3}}r_1r_2e^{-i\varphi }\right) \left| 1\right\rangle
\left\langle 0\right|
\end{eqnarray}
The relation $\det \left( \sigma_A -I \right) =0$ requires that
\begin{equation}
r_2^2=\frac{3\left( 1+\sin ^2\varphi \right) \pm 3\cdot \sqrt{\sin ^4\varphi
-\sin ^2\varphi }}{2+6\sin ^2\varphi } 
\end{equation}
Since $r_2$ is positive, we have
\begin{equation}
\sin ^4\varphi -\sin ^2\varphi \geq 0 
\end{equation}
i.e., 
\begin{equation}
\sin ^2\varphi =1 
\end{equation}
Here another solution $\sin ^2\varphi =0$ is not proper since $0\leq r_2\leq 1$.
So we get 
\begin{equation}
\varphi =\pm \frac \pi 2 
\end{equation}
and 
\begin{eqnarray*}
r_2 &=&\frac{\sqrt{3}}2 \\
r_1 &=&\frac 12
\end{eqnarray*}
Therefore we get two (and only two) different vector solutions:
\begin{eqnarray*}
\overrightarrow{y}^{\left( 1 \right)} &=&\left( \frac 12,\frac{\sqrt{3}}2i\right) \\
\overrightarrow{y}^{\left( 2 \right)} &=&\left( \frac 12,-\frac{\sqrt{3}}2i\right)
\end{eqnarray*}
In order to make the matrix 
\begin{equation}
M=\left( 
\begin{array}{cc}
\sqrt{p} & \sqrt{p}i \\ 
\sqrt{1-p} & -\sqrt{1-p}i
\end{array}
\right) 
\end{equation}
left-unitary (also unitary in this case), there must be 
\begin{equation}
p_1=p_2=\frac 12 
\end{equation}
And we have 
\begin{eqnarray*}
\left( 
\begin{array}{cc}
\frac 1{\sqrt{2}} & \frac 1{\sqrt{2}}i \\ 
\frac 1{\sqrt{2}} & -\frac 1{\sqrt{2}}i
\end{array}
\right) \left( 
\begin{array}{c}
\frac 12\frac 1{\sqrt{2}}\left( \left| \phi ^{+}\right\rangle -i\left| \psi
^{+}\right\rangle \right) \\ 
\frac{\sqrt{3}}2\frac 1{2\sqrt{3}}\left( -3i\left| 00\right\rangle +i\left|
11\right\rangle +\left| 01\right\rangle +\left| 10\right\rangle \right)
\end{array}
\right)   \nonumber \\
=\left(
\begin{array}{c}
\frac 1{\sqrt{2}}\left| 00\right\rangle \\ 
-\frac 1{\sqrt{2}}\left[ \frac 1{\sqrt{2}}\left( \left| 0\right\rangle
+i\left| 1\right\rangle \right) \right] \left[ \frac 1{\sqrt{2}}\left(
\left| 0\right\rangle +i\left| 1\right\rangle \right) \right]
\end{array}
\right) 
\end{eqnarray*}
That is to say, the bipartite state given in Eq.
(\ref{wue3}) is separable, and it can be rewritten as 
\begin{equation}
\rho _{AB}=\frac 12\left| 0\right\rangle _A\left\langle 0\right| \otimes
\left| 0\right\rangle _B\left\langle 0\right| +\frac 12\left| \alpha
\right\rangle _A\left\langle \alpha \right| \otimes \left| \alpha
\right\rangle _B\left\langle \alpha \right| 
\end{equation}
where $\left| \alpha \right\rangle \equiv \frac 1{\sqrt{2}}\left( \left|
0\right\rangle +i\left| 1\right\rangle \right) $.

(3). Let us look at another example. The state of two qutrit systems is given by 
\begin{eqnarray}
\rho _{AB} &=&\lambda \left[ \frac 1{\sqrt{3}}\left( \left| 00\right\rangle
+\left| 11\right\rangle +\left| 22\right\rangle \right) \right] \left[ \frac
1{\sqrt{3}}\left( \left\langle 00\right| +\left\langle 11\right|
+\left\langle 22\right| \right) \right]  \nonumber \\
&& +\left( 1-\lambda \right) \left[ \frac 1{\sqrt{3}}\left( \left|
01\right\rangle +\left| 12\right\rangle +\left| 20\right\rangle \right)
\right] \nonumber  \\
&&\cdot \left[ \frac 1{\sqrt{3}}\left( \left\langle 01\right| +\left\langle
12\right| +\left\langle 20\right| \right) \right]  \label{wue4}
\end{eqnarray}

Set 
\begin{eqnarray}
\left| \Psi \right\rangle &\equiv& y_1\cdot \frac 1{\sqrt{3}}\left( \left|
00\right\rangle +\left| 11\right\rangle +\left| 22\right\rangle \right)
\nonumber   \\
&&+y_2\cdot \frac 1{\sqrt{3}}\left( \left| 01\right\rangle +\left|
12\right\rangle +\left| 20\right\rangle \right) 
\end{eqnarray}
As before, denote $y_1=r_1$, $y_2=r_2\cdot
e^{i\varphi }$, here $r_1$, $r_2$ are positive numbers satisfying the
relation $r_1^2+r_2^2=1$, and $\varphi $ is a real number.

Direct calculation gives
\begin{equation}
\sigma _A=\frac 13\left( 
\begin{array}{ccc}
1 & r_1r_2e^{-i\varphi } & r_1r_2 \\ 
r_1r_2 & 1 & r_1r_2e^{-i\varphi } \\ 
r_1r_2e^{-i\varphi } & r_1r_2 & 1
\end{array}
\right) 
\end{equation}
The condition $\det \left( \sigma_A -I \right) =0$ requires
\begin{equation}
6r_1^2r_2^2e^{-i\varphi } + r_1^3r_2^3\left( 1+e^{-3i\varphi }\right) =8
\end{equation}
which is obviously impossible since $r_1 r_2 \leq \frac {1} {\sqrt{2}}$.
So there is no solution of Eqs. (\ref{wuth}).

Thus we conclude that the bipartite qutrit state given by Eq. (\ref{wue4})
is always entangled. In this example, the same result will be obtained if we
use the rank separability criterion derived by Horodecki et al. \cite{hstt},
since the total density matrix has rank 2 while the reduced density matrices
have ranks 3. 

In conclusion, we have provided a necessary and sufficient condition for
the separability of multipartite states. The key procedure of our criterion
is to solve a set of equations, these equations can be solved
analytically if the density matrix of the given multipartite
state has few nonzero eigenvalues, while numerical approach is always
possible, in this sense, our criterion is operational.

The authors would like to thank Dr. Guang Hou, Jindong Zhou, Prof. Qiang Wu,
Minxin Huang, Yifan Luo, Ganjun Zhu, Guojun Zhu, Jie Yang for helpful
discussions. This project is supported by National Natural Science Foundation
of China under Grant No. 19975043.

%\widetext

\end{document}